\documentclass[%
 reprint,
 amsmath,amssymb,
 aps,
prb,
]{revtex4-1}

\usepackage[pdftex]{graphicx}
\usepackage{dcolumn}
\usepackage{bm}
\usepackage{float}

\begin{document}

\preprint{APS/123-QED}

\setlength{\abovecaptionskip}{0pt}

\title{Nonlinear Pauli Susceptibilities in Sr$_3$Ru$_2$O$_7$ and Universal Features \\ of Itinerant Metamagnetism }

\author{B. S. Shivaram, Jing Luo, and Gia-Wei Chern}

\affiliation{Department of Physics, University of Virginia, Charlottesville, VA. 22901, USA.}

\author{Daniel Phelan}

\affiliation{Materials Science Division, Argonne National Labs, Lemont, IL. 60637.}

\author{R. Fittipaldi and A. Vecchione}

\affiliation{CNR-SPIN, Fisciano, Salerno I-84084, Italy\\ and \\ Dipartimento di Fisica ``'E.R. Caianiello'', Università di Salerno, Fisciano, Salerno I-84084, Italy}

\date{\today}

\begin{abstract}
We report, for the first time, measurements of the third order, $\chi_3$ and fifth order, $\chi_5$, susceptibilities in an itinerant oxide metamagnet, Sr$_3$Ru$_2$O$_7$ for magnetic fields both parallel and perpendicular to the c-axis.  These susceptibilities exhibit maxima in their temperature dependence such that $T_1 \approx 2T_3 \approx 4T_5$ where the $T_i$ are the position in temperature where a peak in the $i$-th order susceptibility occurs.  These features taken together with the scaling of the critical field with the temperature $T_1$ observed in a diverse variety of itinerant metamagnets find a natural explanation in a single band model with one Van Hove singularity (VHS) and onsite repulsion $U$.   The separation of the VHS from the Fermi energy $\Delta$, sets a single energy scale, which is the primary driver for the observed features of itinerant metamagnetism at low temperatures.

\begin{description}

\item[PACS numbers]

75.30.Mb, 75.20.Hr

\end{description}

\end{abstract}

\pacs{Valid PACS appear here}

\maketitle

Metamagnetism (MM), the sudden rise of the magnetization at a critical field, is a phenomenon observed in a diverse range of metals, and extensively studied in both d and f-electron based itinerant systems.~\cite{GiordanoRev1977}  In many itinerant metamagnets as is the case with heavy fermion materials there is a clear presence of local moments, often antiferromagnetically coupled (as ascertained, for instance, from a high temperature Curie Weiss plot), which however develop a strong net moment at the critical field, $B_c$.  The predominant antiferromagnetic correlations are strongly suppressed in high fields \cite{RossatMignod1988} with possibly new ferromagnetic correlations arising \cite{Sato2004} in the vicinity of $B_c$.  In contrast, itinerant electron MM can also be found in systems where there is no clear evidence for local moments \cite{Riveroarxiv} as in the case of the metallic oxide Sr$_3$Ru$_2$O$_7$ (SRO).  Given this fundamental distinction between the two cases it would be natural to ask what if any universal features and/or significant differences exist in the nature of their metamagnetism.

The bilayer ruthenate SRO shows a complex phase diagram where multiple MM transitions may be tuned by varying the angle of the applied field with respect to the crystal axes. Upon increasing temperature, these first-order MM transitions end at critical end points, which are themselves connected by a line of second order phase transitions.~\cite{Lester15,Tokiwa16} Enclosed between these transition lines is an anomalous phase with unusual transport properties. The resistivity in this regime is anomalously high and shows anisotropy. Earlier attempts to understand these anomalous behaviors have been focused on the emergence of a nematic phase associated with the MM transitions.~\cite{Kee2005, Yamase07,Raghu2009} It has also been proposed that the anomalous phase can be viewed as a magnetic analogue of the spatially inhomogeneous superconducting Fulde Ferrell Larkin Ovchinnikov state.~\cite{Berridge09,Conduit09} Interestingly, recent magnetic neutron scattering showed that a spin-density-wave phase is induced in this regime of the phase diagram,~\cite{Lester15} which provides a natural explanation for the strong resistivity anisotropy or the electronic nematic behavior.

Although the nature of the anomalous phase remains to be solved, most theoretical models~\cite{Kee2005, Yamase07,Raghu2009,Berridge09,Conduit09} for the MM transitions assume the presence of a Van Hove singularity (VHS) that is proximate to the Fermi surface in SRO. Experimentally, the existence of VHS within a few meV distance from the Fermi level has indeed been observed in recent high resolution ARPES measurements.~\cite{Tamai2008} Indeed, Binz and Sigrist have demonstrated that MM transitions can be produced in a minimum single-band model with a logarithmically divergent VHS close to the Fermi surface.~\cite{Binz2004} Incorporating weak local Coulomb repulsion between the electrons, they showed that when the magnetic field tunes the Fermi surface of one spin species close enough to the VHS, there is a jump in magnetization. The VHS also plays an important role in models of other metamagnetic materials,~\cite{Evans1992,Kusminskiy08,Beach08} such as the Kondo-lattice model for heavy fermions. For example, divergent density of states (DOS) at the edge of the so-called hybridization gap is a generic feature of quasi-particle band structures in heavy fermions.~\cite{Coleman}

\begin{figure}
\includegraphics[width=1.\columnwidth]{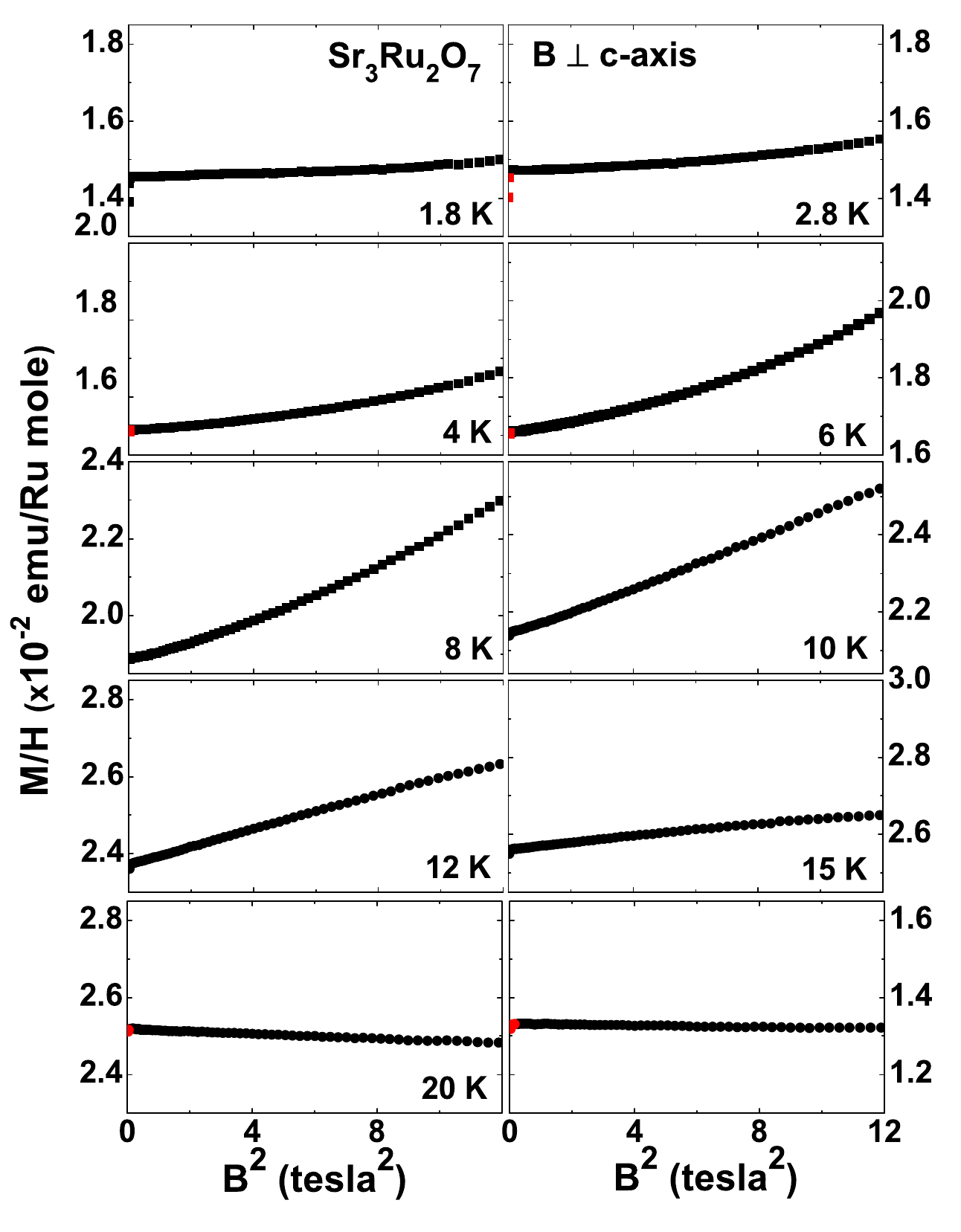}
\caption{\label{fig.1} The magnetization isotherms for $B$ perpendicular to the c-axis plotted as per Eq.~(1) for various temperatures as indicated.  Since the slope yields $\chi_3$ in such a plot it is immediately seen that $\chi_3$ is weakly negative at high temperatures, $T>20$ K, turns positive as $T$ is lowered reaching a maximum around 10 K and decreasing thereafter.  Also noticeable between 3 K and 8 K is the positive curvature which implies a non-zero (+ve) value for the next higher susceptibility, $\chi_5$.}
\end{figure}

In this paper, we present new measurements of nonlinear susceptibilities in high quality single-crystals of SRO for magnetic fields both parallel and perpendicular to the c-axis of the crystal. The experimental results bear a surprising resemblance to the recent work reported by us on heavy fermions and the universal behavior noted there.~\cite{ShivaramUniversality2014, ShivaramChi52014, ShivaramUltrasound2015} We further show that the universal behavior of the nonlinear susceptibilities is a generic feature of Pauli paramagnetism for electronic systems whose Fermi level lies close to a VHS. 

\begin{figure}
\includegraphics[width=1.\columnwidth]{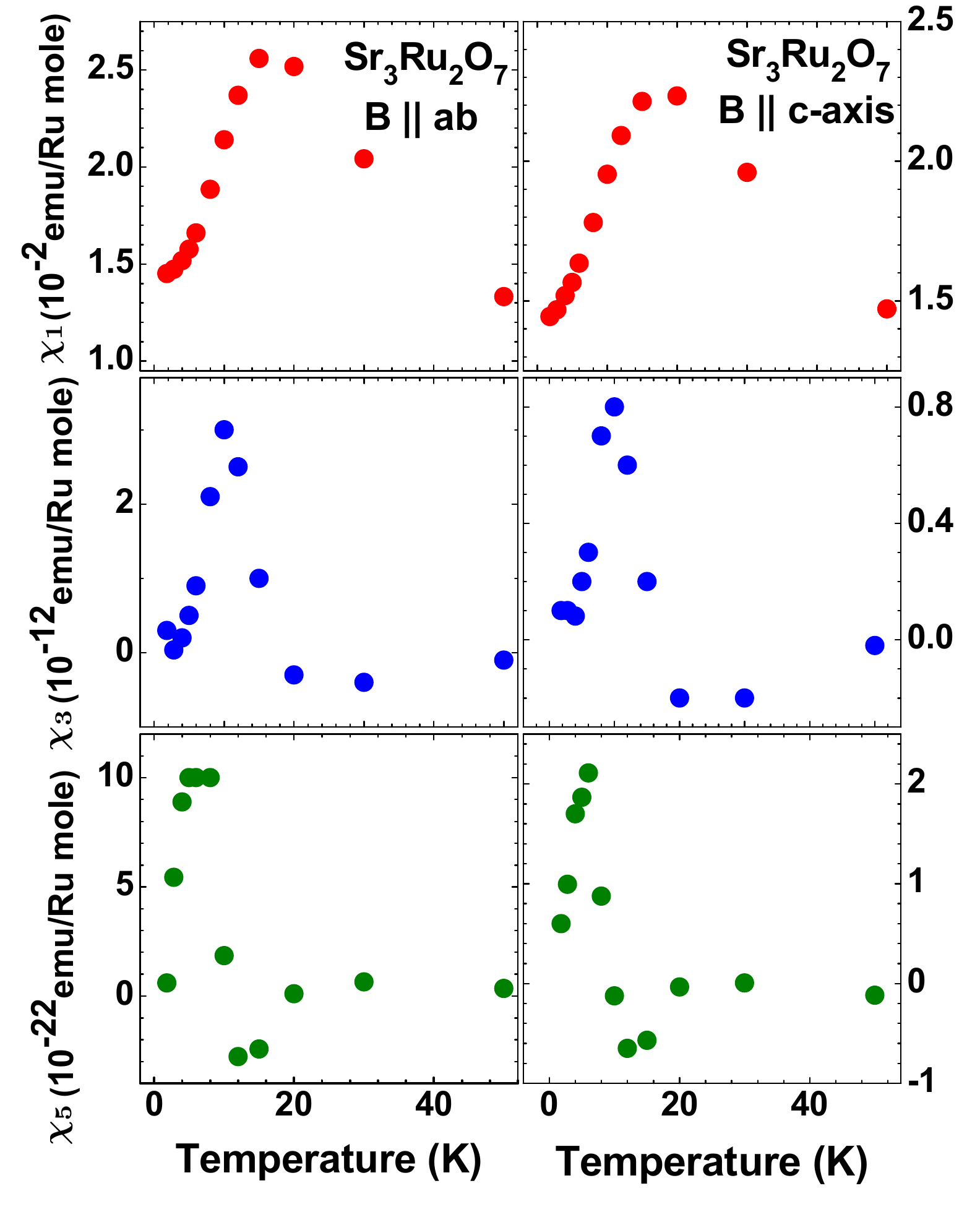}
\caption{\label{fig.2}The values of $\chi_1, \chi_3$ and $\chi_5$ obtained from quadratic fits to the lines such as those shown in Fig.~1.  Note the distinct peaks in all three of these susceptibilities with each higher order susceptibility exhibiting a positive peak at temperatures successively lower by approximately a factor of two. }
\end{figure}


For our study we used single crystals synthesized with a flux growth technique at the University of Salerno. \cite{Ciancio2009}  Their quality was checked by X-ray rocking curves and only the best samples were selected for the present work.  Measurements of the DC magnetization were carried out at Argonne in a commercial SQUID magnetometer (Quantum Design MPMS-3).  The magnetization isotherms obtained at the lowest temperature for $B$ perpendicular to c-axis exhibit apart from the large jump in the magnetization at 5 T a smaller feature at 5.8 T thus confirming the high quality of our samples in accordance with previous work. \cite{Ohmichi2003, Perry2004}

In Fig.1 we show the experimental magnetization isotherms plotted in a particular manner so that the extraction of the nonlinear susceptibilities is facilitated.  The equilibrium magnetization may be written as an expansion in odd powers of the applied field $B$ as:
\begin{equation}	
	M={\chi }_1B+{\chi }_3B^3+{\chi }_5B^5 
\end{equation} 
Dividing both sides of Eq.~(1) by $B$ indicates that a plot of $M/B$ vs $B^2$ yields a straight line with the intercept giving the linear susceptibility and the slope yielding the leading nonlinear susceptibility $\chi_3$ (provided that $\chi_5$ is negligible).  A significant non-zero value of $\chi_5$ would show up as a curvature in the lines in such a plot.  It is indeed observed in Fig.~1 that the lines have a negative slope at the high temperature end, hence negative $\chi_3$ as might be expected in any paramagnet.  However, the slope turns positive as the temperature is lowered and goes through a maximum at a temperature $T_3 \approx 10 K$.  It is well known through several previous measurements \cite{Ikeda2000} that the linear susceptibility in SRO has a maximum at a temperature $T_1 =18$ K.  We thus observe that $T_3 \approx 0.5 T_1$.  Fig.~1 also demonstrates that the next higher order susceptibility $\chi_5$ which is negligible at $T > 10$ K is non-zero for $T < 10$ K where it has a significant positive value.  At the lowest temperature measured $T=1.8$ K the value of $\chi_5$ is nearly zero again.  Thus $\chi_5$ also goes through a maximum albeit at an even lower temperature labelled  $T_5 \approx 0.25 T_1$.  Fig.~2 presents all the three susceptibilities extracted from plots such as those in Fig.~1 for both orientations of the magnetic field.  The raw data plots for the parallel orientation are shown in the supplementary section.

Apart from the discussion above several additional points about the behavior of the three susceptibilities shown in Fig.~2 are noteworthy.  The linear susceptibility has a large non-zero value as $T \to 0$ as it should be in a Pauli enhanced paramagnet.~\cite{Cava1995, Cava2004}   This is true for both $\chi_3$ and $\chi_5$ as $T  \to  0$ for $B_c$.  But the third order susceptibility appears to approach zero at low T.  In addition, while the peak values of the linear susceptibilities are nearly the same, the peak values of the nonlinear susceptibilities are significantly different between the two orientations.  A magnetic field in the basal plane appears to generate a higher nonlinearity as measured by $\chi_3$ and $\chi_5$.  

Experimental results analogous to the above in a completely different family of materials, the f-electron based heavy fermion systems, were considered recently using a simple phenomenological model, which can be thought of as an effective spin-1 system with a large anisotropy term at the single site level.~\cite{ShivaramUniversality2014, PradeepLongPaper} This model reproduced with a remarkable degree of success the experimental correlations such as $T_5 \approx 1/2T_3 \approx 1/4T_1$, the high field magnetic response as well as the ultrasound velocity measurements \cite{ShivaramUltrasound2015}.   While the same model could also be applied to our present work, we note however that this model produced all susceptibilities tending to zero as $T \to$ 0, contrary to the observations here as well as with the heavy fermions.  

In an attempt to further understand results on the nonlinear susceptibilities, here we investigate the minimum theoretical model that includes a VHS proximate to the Fermi edge and a local Hubbard repulsion $U$.~\cite{Binz2004,Berridge11} In the Hartree-Fock mean-field approximation, the Gibbs free energy of the system is given by 
\begin{eqnarray}
	\mathcal{F} &=& -T \sum_{\sigma = \uparrow,\downarrow}  \int d\varepsilon\, \rho(\varepsilon) \ln\left(1 + e^{-\beta(\varepsilon - \mu_\sigma)}\right) \nonumber \\
	& &  + (\mu_\uparrow n_\uparrow + \mu_\downarrow n_\downarrow) +  U n_\uparrow n_\downarrow - B m,
\end{eqnarray}
where $m$ is the normalized magnetization along the field direction, $n_{\uparrow, \downarrow} = n/2 \pm m$ are the densities of up- and down-spin electrons, respectively, $\mu_{\uparrow, \downarrow}$ are the corresponding chemical potentials,  and $\rho(\varepsilon)$ is DOS. Specifically, we consider a DOS with a logarithmically divergent VHS: $\rho(\varepsilon) = (1/W) \ln |W/(\varepsilon - \varepsilon_{\rm VHS})|$, where the parameter $W$ is of the order of the bandwidth. An important term in the model is $\Delta \equiv \varepsilon_{\rm VHS} - \mu$ which controls the ``distance'' of the Fermi level to the singularity. For a given external field $B$, the magnetization is determined from the minimization $\partial \mathcal{F} / \partial m = 0$ subject to the condition that $n_\uparrow + n_\downarrow = n$.~\cite{Binz2004,Berridge11,WohlfarthRhoades1962} More details can be found in the supplemental information.

\begin{figure}
\includegraphics[width=0.95\columnwidth]{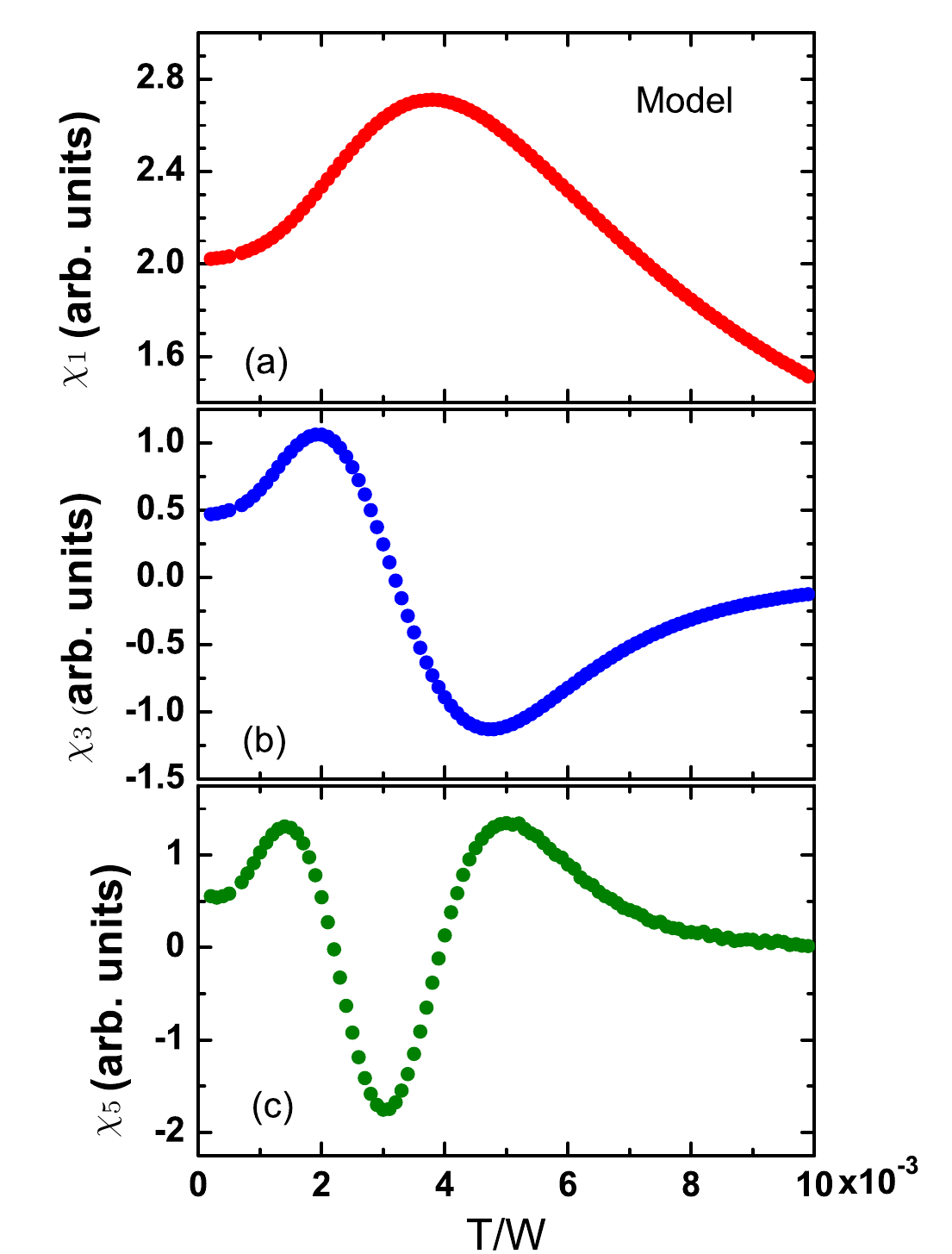}
\caption{\label{fig.3}Shows the calculated susceptibilities  ${\chi }_1,{\chi }_3,{\chi }_5$ in the model.  For the results shown the difference between the chemical potential at $T=0$ and the van Hove singularity $\Delta/W =({\varepsilon }_{\rm VHS}-\mu)/W =0.0095$,  and $U/W=0.19$. }
\end{figure}


Our numerical calculations find a number of remarkable correlations relevant to the interpretation of our experiments.   Fig.~3 shows the calculated susceptibilities from this model. The resemblance to the experimental results of Fig.~2 is striking -- the maxima in all the three susceptibilities are reproduced fairly well. Moreover, we have checked that similar curves are obtained with other types of singularities, e.g. a power-law divergent DOS: $\rho(\varepsilon) \sim 1/|\varepsilon - \varepsilon_c|^\alpha$.
We emphasize that our calculation is based on a minimum yet very general theoretical model which assumes a local Hubbard repulsion and a VHS in the DOS; the model is characterized only by two dimensionless parameters $\Delta/W$ and $U/W$. Consequently,  the results summarized in Fig.~4 represent generic behavior of Pauli linear and nonlinear susceptibilities of electron systems whose Fermi level lies close to a VHS. 

It is worth noting that nonlinear susceptibility measurements can offer unique information about the magnetic properties of materials. For example, a negatively divergent $\chi_3(T)$ provides a direct signature of spin-glass transitions.~\cite{Chalupa77,Suzuki77} It has also been widely used as a probe of quadrupolar spin fluctuations in rare-earth and heavy fermion compounds.~\cite{Morin81,Ramirez92,Bauer06} However, nonlinear susceptibilities have not been systematically investigated in the context of Pauli paramagnetism within a mean-field treatment of Hubbard interaction. Our calculation here thus provides a tell-tale signature of nonlinear Pauli susceptibilities when the Fermi surface is close to a VHS. In particular, the temperatures $T_i$ at which the maximum $\chi_i$ occur display a universal relation $T_5 \approx 1/2T_3 \approx 1/4T_1$, which indeed is observed in many itinerant metamagnets.

In parts (a) and (b) of Fig.~4, we show the ratio $T_1/T_3$ computed for a range of values of the parameters $\Delta$ and $U$.   Remarkably we find that the values for this ratio cluster in the range 2.0 to 2.4 over a wide range of $\Delta$ (i.e. for  $\Delta >0.01 W$) irrespective of the value of $U$ and over a large range of $U< 0.15 W$ irrespective of the value of $\Delta$. As shown in Fig.~4(c), very significantly we find  that the peak temperature $T_1$, which depends linearly on $\Delta$, does {\em not} depend on the value of $U$.  
Furthermore for a noticeable peak to appear in the temperature dependence of the linear susceptibility, the ratio $\chi_1(T_1)/\chi_1(0)$ must be larger than unity.  And as shown in Fig.~4(d), this occurs for a value of $U > 0.1 W$.  Given the experimental value of $\chi_1(T_1)/\chi_1(0) \approx 1.7$ this indicates a Hubbard repulsion $U \approx 0.18 W$ for SRO. This value is consistent with those inferred in previous studies, indicating that SRO is a moderately correlated system.

\begin{figure}[t]
\includegraphics[width=1.\columnwidth]{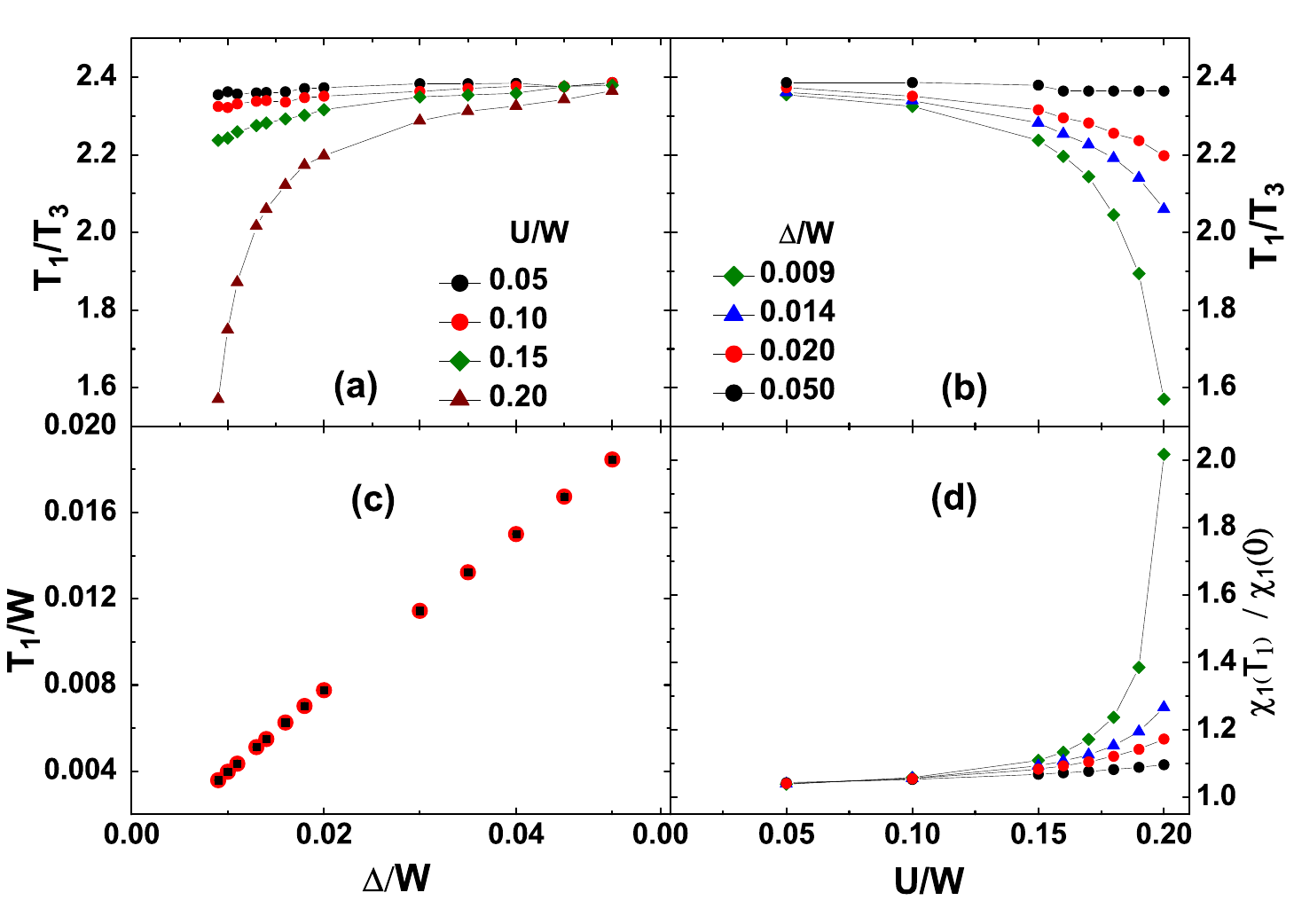}
\caption{\label{fig.4} Parts (a) and (b) show the ratio $T_1/T_3$ computed in the model for a range of values $\Delta$ for specific values of $U$ (left panel) and for a range of values of $U$ for specific values of $\Delta$ (right panel).  Except for very small $\Delta/W \lesssim 0.01$ and very large $ U/W \gtrsim 0.15$ the values of this ratio cluster in the range between 2.4 and 2.0. Part (c) demonstrates that $T_1$ is independent of $U$ but depends only on $\Delta$.  Part (d) shows the ratio $\chi_1(T_1)/\chi_1(0)$ as a function of $U$ for four different values of $\Delta$ as indicated.
}
\end{figure}

These theoretical results prompt us to consider to what extent the current model extends to all itinerant MMs.  We have already noted the ratio $T_3/T_1$ approximates 1/2 in heavy fermion systems.  The empirical correlation of $T_1$ and $B_c$ is a phenomenon that has been even more widely established (Fig.~5).  As stated earlier these correlations were captured in a simple ``local" $S=1$ model.~\cite{ShivaramUniversality2014}  It is remarkable that in the present model where we calculate only the Pauli part, albeit with a specific band structure feature, we find exactly the same correlations as in the ``local" model employed earlier.  Such correlations are also reproduced in an infinite range model for clusters of spins worked out recently by Kumar and Wagner.~\cite{KumarArxiv}  These are unexpected and remarkable coincidences and could very well explain why the diverse set of materials, with varying crystal structures and belonging to different d and f-electron systems (Fig.~5) exhibit the same universal features.  From the success of the present theoretical work it appears that a common factor in all these materials is the occurrence of a Van Hove type singularity and its proximity to the Fermi edge.

It is natural to ask whether experimentally there is evidence for the existence of such singularities uniformly across all systems.  As discussed above, recent high resolution ARPES measurements ~\cite{Tamai2008} in SRO have shown that such a singularity in close proximity (within a few meV) to the Fermi edge indeed exists.  Its observation in heavy fermion systems has also been noted in several systems.  In particular in CeCoGe$_2$ the Kondo temperature is very large, $\sim$250 K, and this facilitates the successful identification/separation of a VHS/Kondo type resonance.~\cite{Im2008, Mun2004}  Identifying similar features in the vast majority of the compounds referred to in Fig.~5 is an experimental task worth undertaking.

\begin{figure}[t]
\includegraphics[width=1.\columnwidth]{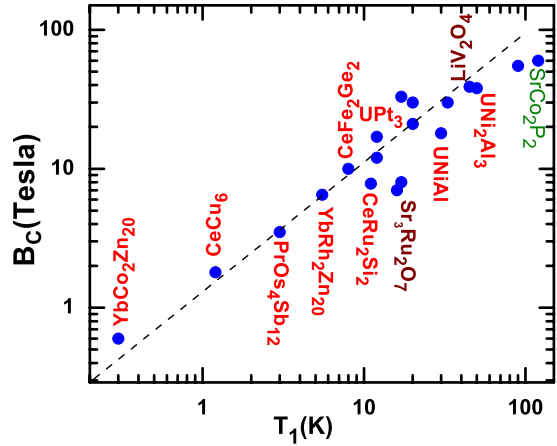}
\caption{\label{fig.5}Shows the linear correlation of $T_1$ vs $B_c$ in a wide variety of heavy fermions, oxides, and pnictides.}
\end{figure}

It is clear from the above discussion and the new experimental results on SRO that a simple microscopic model with a single energy scale albeit with a specific band structure feature is successful to a large extent in providing a full explanation of the linear and nonlinear magnetic response of this strongly correlated itinerant MM.  It is also remarkable that the observed behavior in a system with no evidence for local moments bears a strong resemblance to our earlier work on heavy fermions. Other correlations related in general to the thermodynamics of MMs originating from the current model will be presented in a forthcoming longer paper.

\medskip

\medskip

\textbf{Acknowledgement.} We acknowledge many enlightening conversations with Vittorio Celli and Pradeep Kumar and thank them for a careful reading of the manuscript.  We also thank Geetha Balakrishnan for useful guidance. The magnetometry was sup­ported by the U.S. De­part­ment of En­ergy, Of­fice of Sci­ence, Ba­sic En­ergy Sci­ences, Ma­te­ri­als Sci­ence and En­gi­neer­ing Di­vi­sion.  The magnetometry work at Argonne was supported by the US Department of Energy, Office of Science, Basic Energy Sciences, Materials Science and Engineering Division. \\

\renewcommand{\thefigure}{S\arabic{figure}}
\setcounter{figure}{0}

\bigskip

{\bf \large \centering Supplemental Materials \par}

\appendix

\section{Model}
\label{sec:model}

In this section, we present details of the nonlinear susceptibility calculation. We consider a minimum theoretical single-band model with a local Hubbard repulsion, described by the following Hamiltonian
\begin{eqnarray}
	\mathcal{H} = \sum_{\mathbf k, \sigma= \uparrow, \downarrow} (\varepsilon_{\mathbf k} - \mu - \sigma B) c^\dagger_{\mathbf k, \sigma} c^{\;}_{\mathbf k, \sigma} + U \sum_i n_{i, \uparrow} n_{i, \downarrow} \quad
\end{eqnarray}
where $\varepsilon_{\mathbf k}$ denotes the band energy of electrons, $\mu$ is the Fermi level, $B$ is the external magnetic field, and $U$ is the Hubbard parameter. The density of states (DOS) is given by the following summation
\begin{eqnarray}
	\rho(\varepsilon) = \frac{1}{V} \sum_{\mathbf k} \delta(\varepsilon - \varepsilon_{\mathbf k})
\end{eqnarray}
Here we employ the Hartree-Fock mean-field (MF) approximation to compute the itinerant magnetization of the electrons. Since the only information about the band-structure which enters the MF calculation is the DOS, our calculation focuses on how the presence of a Van~Hove singularity (VHS) in $\rho(\varepsilon)$ affects the Pauli paramagnetism. 

For a given $\rho(\varepsilon)$, the electron density is expressed as
\begin{eqnarray}
	n_\sigma = \int d\varepsilon \rho(\varepsilon) \, f(\varepsilon - \mu_\sigma)
\end{eqnarray}
where $f(\varepsilon) = 1/(\exp(\beta\epsilon) + 1)$ is the Fermi-Dirac function, $\beta = 1/k_B T$ is the inverse temperature, and $\mu_\sigma$ denotes the chemical potential of electrons with spin $\sigma = \uparrow, \downarrow$. These two chemical potentials are determined from the minimization of the Gibbs free energy subject to the constraint $n_\uparrow + n_\downarrow = n$, which is the filling fraction. To proceed, we first introduce the magnetization $m$ such that $n_{\uparrow, \downarrow} = \frac{n}{2} \pm m$. The Gibbs free energy density is then written as:
\begin{eqnarray}
	\mathcal{F} &=& -T \sum_{\sigma = \uparrow,\downarrow}  \int d\varepsilon\, \rho(\varepsilon) \ln\left(1 + e^{-\beta(\varepsilon - \mu_\sigma)}\right) \nonumber \\
	& &  + (\mu_\uparrow n_\uparrow + \mu_\downarrow n_\downarrow) +  U n_\uparrow n_\downarrow - B m
\end{eqnarray}  
Minimization of $\mathcal{F}$ with respect to $m$ yields the following self-consistent equation ~\cite{SIBinz2004}
\begin{eqnarray}\label{BUm}
	B = \mu_\uparrow(n, m) - \mu_\downarrow(n, m) - 2 U m
\end{eqnarray}
from which we can obtain implicitly the field dependence of electron magnetization. 

Following Ref.~\onlinecite{SIBerridge11} and \onlinecite{SIBerridge11Jphys}, we consider a logarithmically divergent VHS for the DOS:
\begin{eqnarray}
	\rho(\varepsilon) = \frac{1}{W} \ln \left|\frac{W}{\varepsilon - \varepsilon_{\rm VHS}} \right|
\end{eqnarray}
Here $W$ sets the energy scale of the model and also serves as a measure of the bandwidth. Another important parameter is $\Delta = \varepsilon_{\rm VHS} - \mu$, which controls the distance between the VHS and the Fermi edge.

We compute the linear and nonlinear susceptibilities ${\chi }_1,{\chi }_3,{\chi }_5$ following exactly the same procedure as the experimental one. Explicitly, the above mean-field calculation is repeated for varying magnetic field to obtain a magnetization curve $m(B)$. Expanding $m$ as a power series of $B$ (only the odd powers enter the expansion due to time-reversal symmetry), we have
\begin{eqnarray}
	\label{eq:m-expansion}
	m = \chi_1 B + \chi_3 B^3 + \chi_5 B^5 + \cdots
\end{eqnarray}
By numerical values of $\chi_i$ are obtained via fitting the $m/B$ vs $B^2$ curves: the intercept gives the linear susceptibility $\chi_1$, the slope yields $\chi_3$, and the curvature $\chi_5$. The temperature dependence of the extracted susceptibilities is shown in Fig.~3 of the main text. In this calculation, we have used parameters $W=10$ (which sets the energy scale), $\Delta/W = 0.0095$, and $U/W = 0.19$.

Our calculation shows that the ratio $\chi_1(T=T_1)/\chi_1(T=0)$ is very sensitive to the Hubbard parameter $U$. This ratio grows with increasing $U$ as shown in Fig.~4(d) of the main text. It also becomes larger with smaller $\Delta$, i.e. when the Fermi energy is close to the VHS. The experimentally observed ratio $\chi_1(T=T_1)/\chi_1(T=0) \approx 1.7$ corresponds to a small Hubbard $U \approx 0.18\,W$ relative to the bandwidth, thus justifying the mean-field approach.

As shown in Fig.~4 in the main text, the value of $T_1/T_3$ lies in the interval between 2 and 2.4 as long as the interactions are weak and the VHS is not located too close to the fermi energy. We also find that $T_1$ is proportional to $\Delta$ (when $\Delta$ is not too large) and is independent of Hubbard repulsion $U$. The independence of $T_1$ on $U$ can be understood from the analytical expression for the linear susceptibility for small temperature and field:
\begin{eqnarray}\label{chi10}
	\chi_1= - \frac{\int{d \varepsilon \rho \left( \varepsilon \right) f^\prime \left( \varepsilon -\mu \right) }} 
	{{2 \left[ 1 + U \int d \varepsilon \rho \left( \varepsilon \right) f^\prime \left( \varepsilon -\mu \right)\right]}} 
\end{eqnarray}
Numerically, the denominator is a rather smooth function of temperature. The occurrence of a peak at $T_1$ comes from the temperature dependence of the numerator. Consequently, the peak position $T_1$ remains independent of $U$ as long as~$1+U\int{d\varepsilon \rho \left(\varepsilon \right)f^\prime\left(\varepsilon -\mu \right)} > 0$.

\bigskip

\section{Analytical Calculation of Linear and nonlinear susceptibilities}  

Here we provide the analytical approximation for the linear and nonlinear susceptibilities $\chi_1$, $\chi_3$ of our model. For simplicity, we define a function 
\begin{eqnarray}
	F(x)=\int{d\varepsilon \rho \left(\varepsilon \right)f\left(\varepsilon -x\right)}
\end{eqnarray}
which depends implicitly on temperature through the Fermi-Dirac function $f(\varepsilon-x)$. In terms of $F(x)$, the density of electrons with spin-$\sigma$ is $n_\sigma = F(\mu_\sigma)$. For a given total filling fraction $n = n_\uparrow + n_\downarrow$ and magnetization $m = (n_\uparrow - n_\downarrow)/2$, the two chemical potentials $\mu_\uparrow$ and $\mu_\downarrow$ have to be determined self-consistently by the following equations
\begin{eqnarray}
	\label{eq:A1}
	F(\mu_\uparrow) + F(\mu_\downarrow) = n \\
	\label{eq:A2}
	\mu_\uparrow - \mu_\downarrow = B + 2 U m
\end{eqnarray}
In our calculation, we assume a small $B$ and introduce a parameter $x = \mu_\uparrow - \mu$, representing the deviation of the chemical potential of up-spin electron from the one $\mu$ in the absence of field. The chemical potential for down-spin electron is $\mu_\downarrow = \mu + x - B - 2 U m$.  For a filling fraction $n$, the zero-field chemical potential $\mu$ is given by  $\mu = F^{-1}(n/2)$. 
Next, we expand $n_{\uparrow, \downarrow}$ in terms of the small deviation parameter $x$:
\begin{eqnarray}
n_{\uparrow }=F\left({\mu }_{\uparrow }\right)&=&F\left(\mu \right)+F^\prime\left(\mu \right)x+\frac{1}{2}F^{\prime\prime}\left(\mu \right)x^2 \nonumber \\
 &+& \frac{1}{6}F^{\prime\prime\prime}\left(\mu \right)x^3+\cdots
\end{eqnarray}
\medskip
\begin{eqnarray}
n_{\downarrow }=F\left({\mu }_{\downarrow }\right) &=& F\left(\mu \right)+F^\prime\left(\mu \right)(x-B-2Um) \nonumber\\
\notag
&+& \frac{1}{2}F^{\prime\prime}\left(\mu \right){(x-B-2Um)}^2 \\
&+&\frac{1}{6}F^{\prime\prime\prime}\left(\mu \right){(x-B-2Um)}^3+ \cdots
\end{eqnarray}
Plug these expressions into Eqs.~(\ref{eq:A1}) and (\ref{eq:A2}), we have
\begin{eqnarray}
\notag
0&=&n_{\uparrow }+n_{\downarrow }-n=F^\prime\left(\mu \right)(2x-B-2Um) + \\ 
&+&\frac{1}{2}F^{\prime\prime}\left(\mu \right)\left[x^2+{(x-B-2Um)}^2\right]+\cdots \\
\notag
2m&=&n_{\uparrow }-n_{\downarrow }=F^\prime\left(\mu \right)(B+2Um) \\
&+&\frac{1}{2}F^{\prime\prime}\left(\mu \right)\left[x^2-{(x-B-2Um)}^2\right]+\cdots
\end{eqnarray}
To solve these two equations perturbatively, we introduce a Taylor expansion in terms of $B$ for both $m$ and $x$. The expansion coefficients of the magnetization are simply the linear and nonlinear susceptibilities; see Eq.~(\ref{eq:m-expansion}), while those for $x$ are denoted as $c_k$, i.e. 	$x = c_1 B + c_2 B^2 + c_3 B^3 + \cdots$. Substituting these expansion into the above two equations, we obtain
\begin{eqnarray}
c_1&=& \frac{1}{2\left[1-UF^\prime\left(\mu \right)\right]} \\
c_2&=& - \frac{F^{\prime\prime}\left(\mu \right)}{8F^\prime\left(\mu \right)\left[1-UF^\prime\left(\mu \right)\right]^2} \\
c_3 &=& \frac{-3U{\left[F^{\prime\prime}\left(\mu \right)\right]}^2+UF^\prime\left(\mu \right)F^{\prime\prime\prime}\left(\mu \right)}{48F^\prime\left(\mu \right){\left[1-UF^\prime\left(\mu \right)\right]}^4}\\
\label{chi1}
{\chi }_1&=& \frac{F^\prime\left(\mu \right)}{2\left[1-UF^\prime\left(\mu \right)\right]} \\
{\chi }_3 &=& \frac{-3{\left[F^{\prime\prime}\left(\mu \right)\right]}^2+F^\prime\left(\mu \right)F^{\prime\prime\prime}\left(\mu \right)}{48F^\prime\left(\mu \right){\left[1-UF^\prime\left(\mu \right)\right]}^4}
\end{eqnarray}
where Eq. (\ref{chi1}) corresponds to Eq. (\ref{chi10}) in the Sec.~\ref{sec:model}. $\chi_5$~can be derived by the same calculation with higher order Taylor expansions. A Taylor expansion in temperature can be achieved by substituting $F(\mu)$ and its derivatives with Sommerfeld expansion.

\section{Additional experimental figures and data}  

Shown in Fig.~S1 is the raw data for the magnetization for the parallel orientation plotted in a manner as explained in the main text. The values of $\chi_3$ which start out negative on the high tempeature end turn positive as the temperature is lowered and reach a maximum at a temperature of $\sim$10 K.


Finally, we note yet another result but this one being model independent. Since all the three susceptibilities $\chi_1, \chi_3$ and $\chi_5$  are measured in our work it is possible to evaluate the term $\left(3{\chi }^2_3-{\chi }_1{\chi }_5\right)$.  Through a Landau expansion of the free energy it can be shown that the value of this term is constrained to be greater than zero to ensure thermodynamic stability. This however is not the case experimentally for both field orientations as shown in Fig.~S2 (top).  The term dips below zero at approximately 10 K and continues to be negative down to the lowest temperatures measured.  Such a feature is also produced with the theoretical model employed here, Fig.~S2 (bottom).  The breakdown of this stability condition simply implies that other higher order terms are needed in the expansion of the free energy.  \\

\bigskip

\begin{figure}
\includegraphics[width=0.99\columnwidth]{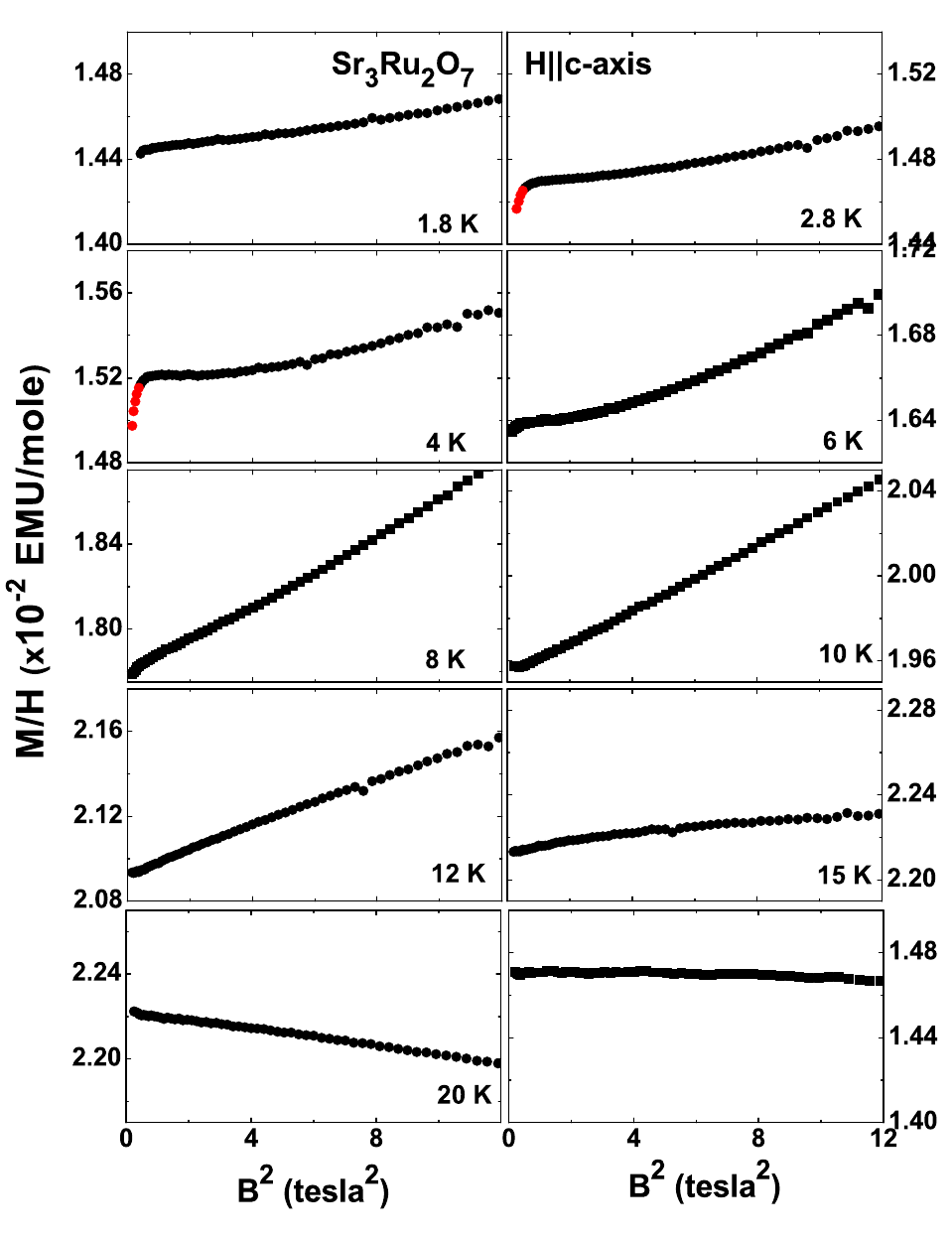}
\caption{\label{fig. S1}Data similar to fig.1 of main text but for the $B \parallel $~c-axis case.}
\end{figure}

\begin{figure}[H]
\includegraphics[width=0.9\columnwidth]{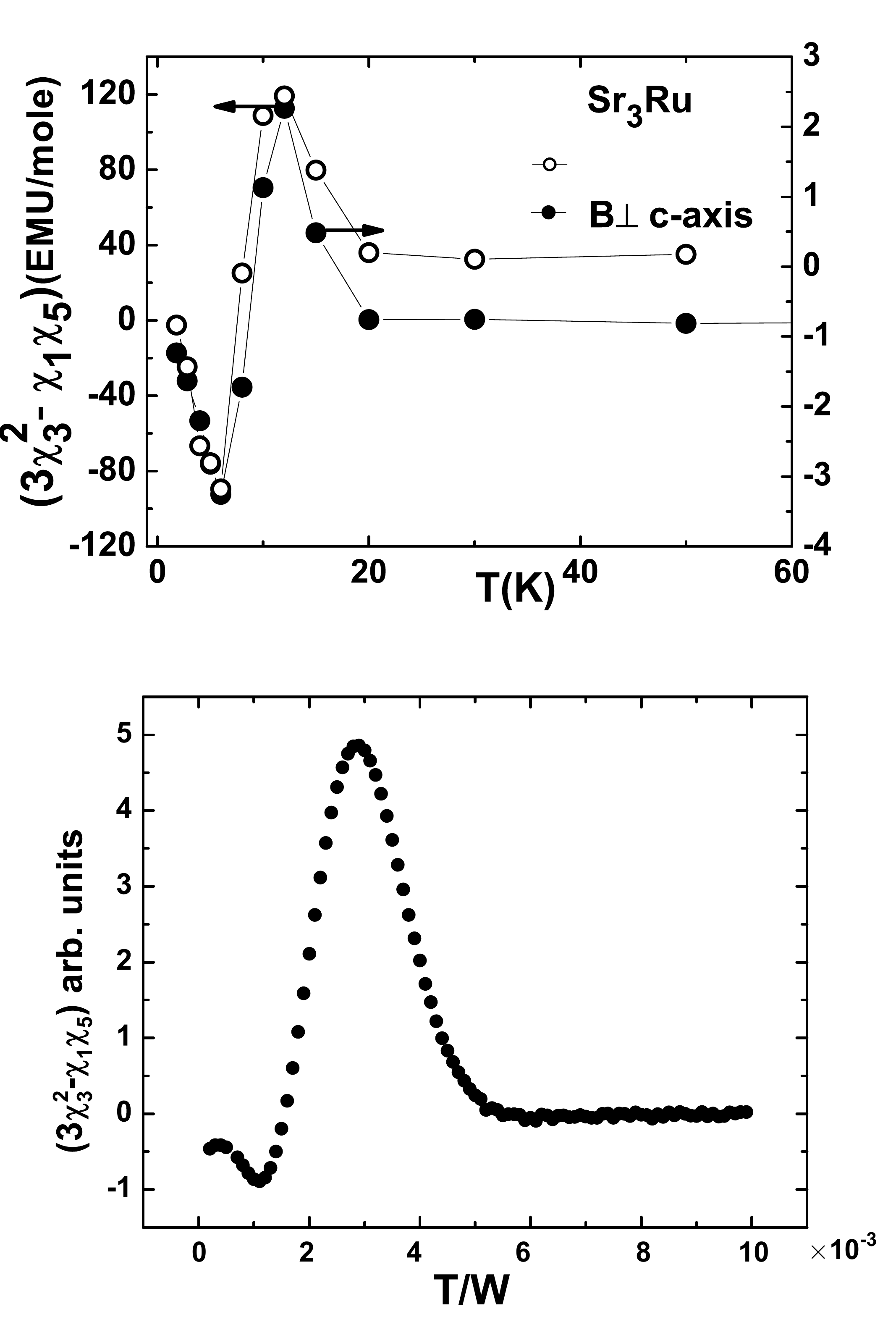}
\caption{\label{fig. S2} (Top) The thermodynamic stability parameter, $(3 \chi_3^2-\chi_1 \chi_5)$, vs temperature for both field orientations. (Bottom) theoretical curve of same parameter vs temperature.}
\end{figure}




\end{document}